\documentclass[aip, rsi, preprint]{revtex4-2}
\usepackage{hyperref}
\usepackage{graphicx}
\usepackage{latexsym}
\usepackage{amsmath,amssymb}
\usepackage{xcolor}
\usepackage[utf8]{inputenc}
\usepackage[english]{babel}

\makeatletter
\def\@email#1#2{%
 \endgroup
 \patchcmd{\titleblock@produce}
  {\frontmatter@RRAPformat}
  {\frontmatter@RRAPformat{\produce@RRAP{*#1\href{mailto:#2}{#2}}}\frontmatter@RRAPformat}
  {}{}
}%
\makeatother

\begin{document}

\preprint{AIP/123-QED}

\title{Tunable biaxial strain device for low dimensional materials.}

\author{Vincent Pasquier}
\author{Alessandro Scarfato}
\author{Jose Martinez-Castro}
\author{Antoine Guipet}
\author{Christoph Renner}

 \affiliation{DQMP, Université de Genève, 24 Quai Ernest Ansermet, CH-1211 Geneva, Switzerland}

\date{\today}

\begin{abstract}

Strain is attracting much interest as a mean to tune the properties of thin exfoliated two-dimensional materials and their heterostructures. Numerous devices to apply tunable uniaxial strain are proposed in the literature, but only few for biaxial strain where there is often a trade-off between maximum strain and uniformity, reversibility and device size. We present a compact device that allows the controlled application of uniform in-plane biaxial strain, with maximum deformation and uniformity comparable to those found in much larger devices. Its performance and strain uniformity over the sample area are modeled using finite element analysis and demonstrated by measuring the response of exfoliated 2H-MoS$_2$ to strain by Raman spectroscopy.

\end{abstract}


\maketitle
\section{INTRODUCTION}
Strain engineering is one of the key strategies to boost the performances of the latest generation transistors. \cite{tsuitsui2019strainreview} Beyond performance, tunable properties are paramount to functional electronic devices. In this context, strain is becoming increasingly fashionable to tune the optical and electronic properties of contemporary materials and devices. Strain is of particular interest in low dimensional materials, starting with graphene \cite{naumis2017strainreview} and transition metal dichalcogenides (TMDs).\cite{roldan2015strainreview, peng2020} The ability to exfoliate these materials into single unit-cell thin crystals and their controlled stacking promote an entirely new approach to material synthesis and device assembly.\cite{butler20132dreview,li20152dreview} Their stretchability makes them particularly susceptible to tune their optical and electromagnetic properties by means of controlled strain.

Strain can be directly embedded into the material or device during its synthesis through an appropriate choice of growth conditions or by combining lattice mismatched materials.\cite{gao2018atomic, zhang2017} However, this approach does not enable any post-fabrication adjustments of the strain. To this end, a number of mechanical devices have been developed to apply tunable strain. They include mounting the sample between movable anvils,\cite{ghini2021, hicks2014strong, edelberg2020tunable} mounting them on a bending \cite{guo2011, michail2020biaxial} or stretching \cite{he2015} beam, or directly onto a piezoelectric crystal.\cite{gao2018atomic} 

\section{BIAXIAL STRAIN DEVICE}

\begin{figure}%
\includegraphics[width=0.6\columnwidth] {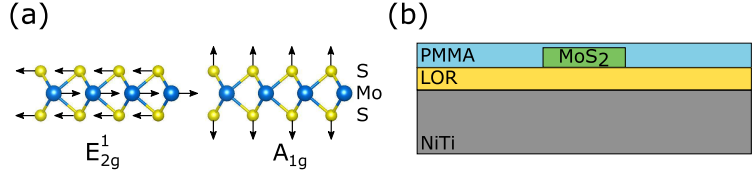}
\caption{(a) Side view of the 2H-MoS$_2$ structure with the active E$^1_{2g}$ in-plane and A$_{1g}$ out-of-plane vibrational Raman modes. (b) Schematic cross-section of a device heterostructure. The base is a Nickel-Titanium alloy plate with a lift off resist (LOR) layer on top. A MoS$_2$ flake is exfoliated onto the LOR and clamped to the substrate via a thin polymer (PMMA) layer.}
\label{sample}
\end{figure}

We developed a device to apply tunable biaxial strain in a confined space by the controlled bending of a thin Nitinol substrate, a superelastic alloy made of $50\%$ nickel and $50\%$ titanium. A suitably smooth surface with 1-2 nm peak-to-peak roughness to deposit very thin exfoliated crystals is achieved by spin-coating a lift off resist (LOR) onto the Nitinol substrate. The material to be strained is either transferred \cite{dean2010boron} or directly exfoliated onto the LOR (Fig.\ref{sample}). For scanning tunneling microscopy applications, a 2 nm titanium buffer layer and a 5 nm thin gold film are evaporated onto the LOR prior to mounting the exfoliated van der Waals crystals for the purpose of electrically connecting them. 

When straining van der Waals materials, we noticed that the interlayer forces were often too weak to transfer the strain from the bent substrate to the top layer of thick crystals. In order to avoid layer slippage and subsequent strain relaxation, the crystals must be firmly clamped down. For the Raman spectroscopy measurements discussed below, the 2H-MoS$_2$ crystals were encapsulated in a thin polymer (PMMA) layer. For scanning probe applications, the polymer layer is replaced by two thin Au/Ti strips ($\approx$ 30 nm thick) evaporated in-situ at a base pressure of $10^{-8}$ mbar over two opposite edges of the crystal through a shadow mask. 

\begin{figure}%
\includegraphics[width=0.6\columnwidth] {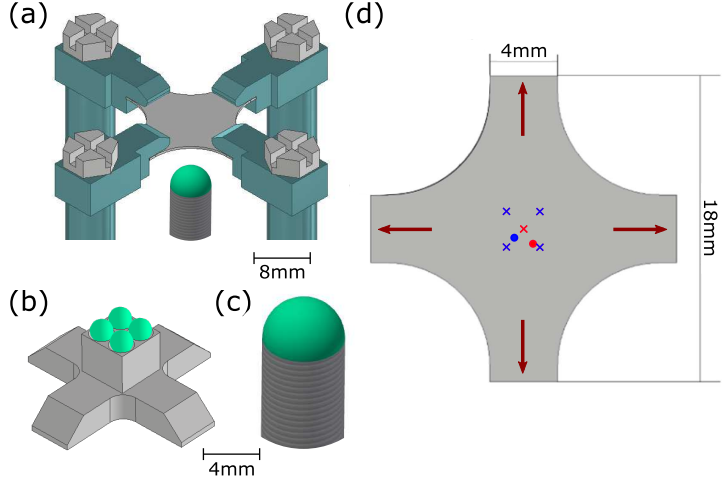}
\caption{Biaxial strain device.  (a) Schematic view of the biaxial strain device with the (b) 4-points and (c) single point pushers. (d) Top view of the substrate bending plate enabling the application of uniform biaxial strain. The red and blue crosses show the application points of the single and four points pushers, respectively. The red and blue dots indicate the positions of the samples during the strain experiments using the single and four points pushers, respectively. The red arrows indicate the direction of strain.}
\label{device}
\end{figure}

The strain in our device is generated by the controlled bending of a Nitinol substrate (Fig.\ref{device}). The bending force is applied by means of a pusher actuated by a screw. All the parts of the device are made of steel, including the top of the pushers made of balls from ball-bearings. The response of the strain device is calibrated using commercial metal strip gauges, \cite{regodic2015development, straingauge} which measure the strain applied along a specific direction from a change in the strip resistance. This calibration is compared to the actual strain applied to the van der Waals material which we infer from changes in characteristic Raman modes of 2H-MoS$_2$.\cite{conley2013bandgap, lloyd2016band}

\begin{figure}%
\includegraphics[width=0.5\columnwidth]{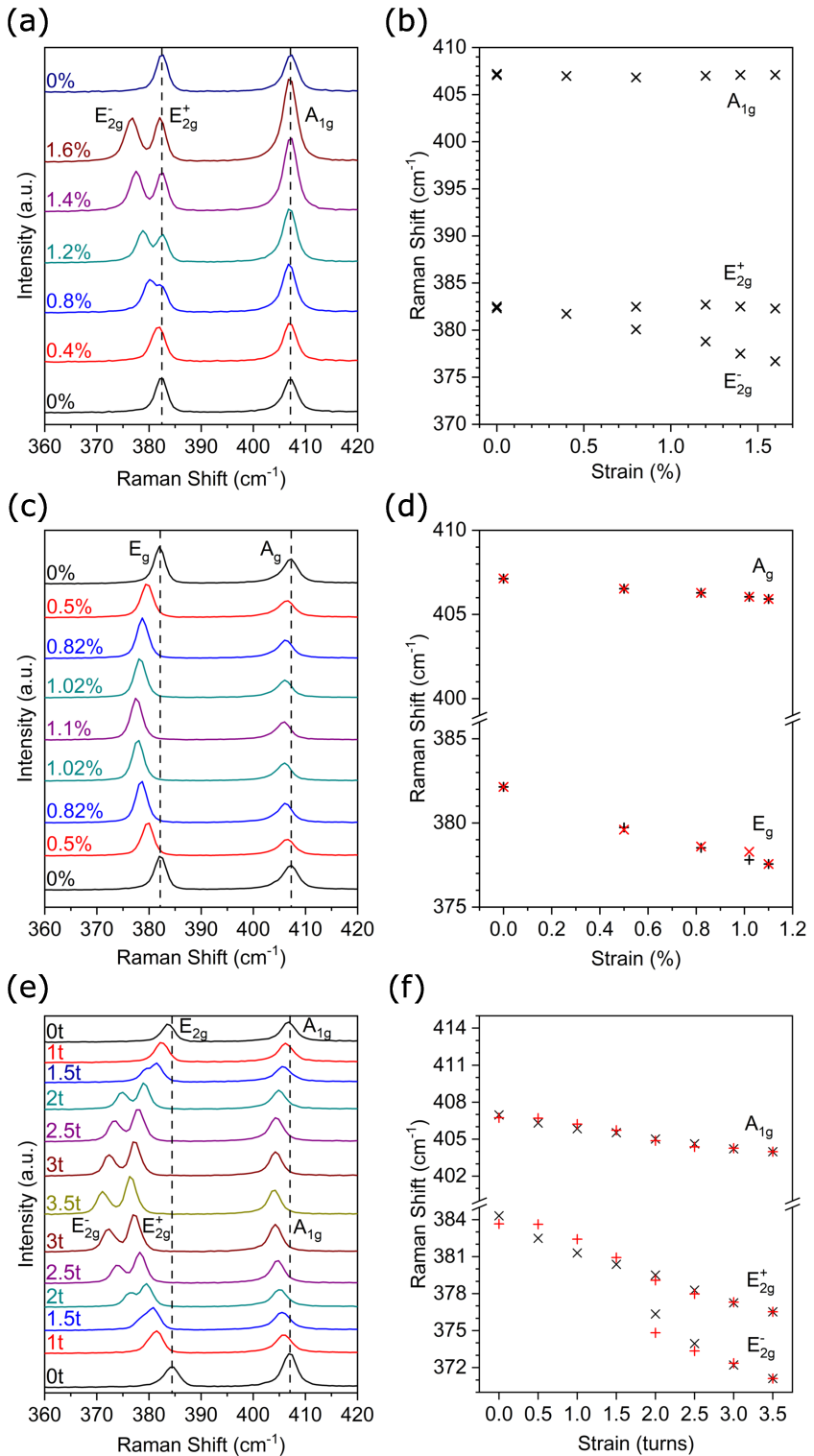}%
\caption{ \label{raman} (left) Raman spectra of 2H-MoS$_2$ as a function of tensile strain - dashed lines indicate the Raman peaks positions in the absence of strain, which are slightly different in the three cases due to different exfoliated sample thicknesses.\cite{li2012bulk} (right) Corresponding in-plane and out of plane mode positions obtained by fitting the Raman peaks to Lorentzians. (a) and (b) correspond to uniaxial tensile strain from 0$\%$ to 1.6$\%$ and back to 0$\%$. (c) and (d) correspond to uniform biaxial in-plane tensile strain from 0$\%$ to 1.1$\%$ and back to 0$\%$. (e) and (f) correspond to non-uniform biaxial in-plane tensile strain from zero to maximal strain and back to zero strain expressed in turns of the pusher screw.}
\end{figure}

\section{RESULTS}

For reference, we first show the Raman response of 2H-MoS$_2$ to uniaxial tensile strain (Fig.\ref{raman}(a)) applied using a standard device based on the bending of a simple rectangular beam.\cite{conley2013bandgap,zhu2013strain} With increasing tensile strain, the out-of-plane A$_{1g}$ mode remains unchanged at 407 cm$^{-1}$ and the in-plane E$_{2g}$ mode splits into E$^-_{2g}$ and E$^+_{2g}$ modes. While the split E$^+_{2g}$ mode stays at around 382.3 cm$^{-1}$, there is a clear red shift of the E$^-_{2g}$ mode at a rate of about 3.5 cm$^{-1}/\%$, in range with previous findings.\cite{conley2013bandgap, zhu2013strain, zhang2021investigation} Upon fully relaxing the strain, the initial unstrained Raman spectrum is recovered, demonstrating a perfectly reversible strain application at least up to 1.6$\%$. Fitting the Raman modes to Lorentzians, we can follow the modes positions as a function of strain as shown in Fig.\ref{raman}(b).

Next, we measure the Raman response of 2H-MoS$_2$ under uniform biaxial tensile strain applied with the new device depicted in Fig.\ref{device}. The corresponding Raman spectra presented in Fig.\ref{raman}(c) are distinctively different from the Raman response to uniaxial strain discussed above. Both the A$_{1g}$ and E$_{2g}$ Raman modes are red shifted with increasing tensile strain, without any splitting of the E$_{2g}$ mode. Contrary to the uniaxial configuration, uniform biaxial tensile strain preserves the in-plane symmetry of the crystal and therefore the degeneracy of the E$_{2g}$ mode. The in-plane mode is more sensitive to strain than the out-of-plane mode. Indeed, from 0$\%$ to 1.1$\%$ strain, the E$_{2g}$ peak shifts from 381.8 cm$^{-1}$ to 377.1 cm$^{-1}$ while the A$_{1g}$ mode shifts from 406.9 cm$^{-1}$ to 405.5 cm$^{-1}$ (Fig.\ref{raman}(d)). This 70$\%$ difference is a direct consequence of a more important in-plane than out-of-plane deformation under in-plane tensile stress. Upon relaxing the strain, the Raman modes recover their unstrained positions, demonstrating the perfect reversibility of the applied biaxial strain. This reversibility is clearly seen in the peak positions as a function of increasing and decreasing strain in Fig.\ref{raman}(c) and (d). Additional strain cycles produce the same results within 0.5$\%$ accuracy.

\section{DISCUSSION}

Numerous devices enabling the application of reproducible uniaxial strain have been described in the literature. \cite{zhang2021investigation, zhu2013strain, conley2013bandgap} The observed splitting of the in-plane Raman mode in Fig.\ref{raman}(a) and (b) reflects the anisotropic lattice distortion due to stress applied along one direction only, which lifts their degeneracy. We emphasize that only properly clamped crystals exhibit the perfectly reproducible response observed in Fig.\ref{raman}(a) and (b) when repeatedly cycling the strain between 0$\%$ and maximum strain. Simply pasting the 2H-MoS$_2$ crystal to the Nitinol substrate without any additional clamping of the surface layer did in many cases not allow to transfer strain to the free surface, resulting in a constant Raman signal despite increasing the applied stress. 

Applying controlled and uniform biaxial strain is significantly more challenging than applying uniaxial strain. We found both the shape of the bending plate and the shape of the pusher to be critical in achieving the sizeable and uniform biaxial in-plane strain demonstrated in Fig.\ref{raman}(c) and (d). Using a simple cross-shaped Nitinol substrate allows to induce only limited strain at the centre of the cross. The central square where the sample is mounted on such a cross is too rigid to be significantly strained by the four rectangular arms. The more elaborate shape shown in Fig.\ref{device}(d) overcomes this limitation and allows sizeable uniform deformation of the central region through bending of the cross (up to 1.4$\%$ demonstrated so far). 

\begin{figure}%
\includegraphics[width=0.6\columnwidth] {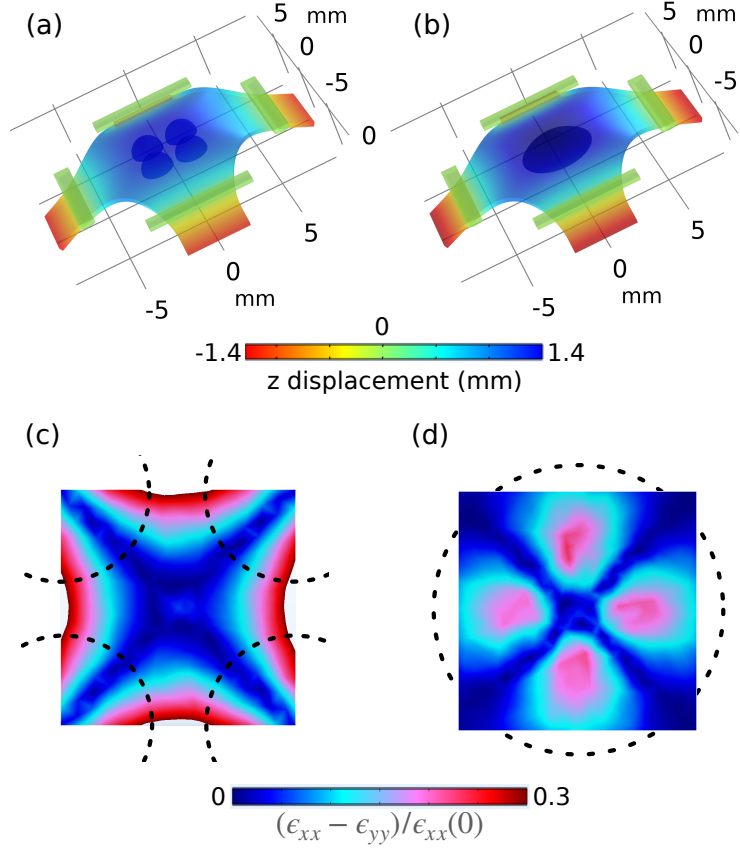}
\caption{Finite element analysis of the Nitinol cross deformation with (a) a four point  pusher and (b) a single point pusher. The color scale corresponds to the cross' displacement along z. Details of the deformation amplitude obtained in the central region of the Nitinol cross are shown in (c) for the four point pusher and in (d) for the single point pusher. The dashed circles depict the position of the steel balls at the top of the pusher. The color scale represents the relative difference of strain between the orthogonal in-plane directions $(\epsilon_{xx}-\epsilon_{yy})/\epsilon_{xx}(0)$, where $\epsilon_{xx}(0)=\epsilon_{yy}(0)$ is the strain at the center of the cross. }
\label{comsol}
\end{figure}

The second key feature of our device is the pusher used to deform the Nitinol cross. First experiments using a simple single point pusher (Fig.\ref{device}(c)) did not yield systematic uniform biaxial strain. Typical Raman spectra obtained in this configuration are shown in Fig.\ref{raman}(e) and (f). Both the A$_{1g}$ and E$_{2g}$ Raman modes shift to higher frequencies as expected for biaxial strain. However, the in-plane E$_{2g}$ mode splits into E$^-_{2g}$ and E$^+_{2g}$ modes with increasing strain, indicating non uniform in-plane stresses. We do not know the precise amount of stress along each orthogonal direction, which is why we label the different Raman curves as a function of the number of turns of the screw moving the pusher rather than in actual strain. Nevertheless, we can estimate the applied stress based on the linear dependence of the Raman peak shift with strain and the uniform strain data presented in Fig.\ref{raman}(c) and (d). The A$_{1g}$ peak shifts by 3 cm$^{-1}$ from 407 cm$^{-1}$ to 404 cm$^{-1}$. Comparing this shift with the 1.4 cm$^{-1}$ shift at 1.1$\%$ uniform strain suggests the applied strain is at least 2.3$\%$. The E$^1_{2g}$ mode shifts from 384.3 cm$^{-1}$ into the split E$^-_{2g}$ and E$^+_{2g}$ modes at 371.1 cm$^{-1}$ and 376.5 cm$^{-1}$ respectively at maximum strain. This shift is again about twice the shift observed under uniform biaxial strain, suggesting a non uniform strain in excess of 2.2$\%$, consistent with the estimate based on the A$_{1g}$ mode. The difference between the strain along the two orthogonal in-plane directions can be estimated from the splitting of the E$^1_{2g}$ mode which corresponds to the splitting of this peak observed at about 1.5$\%$ uniaxial strain, suggesting about 50$\%$ more strain in one direction. This analysis shows that the single point pusher can achieve larger strain than the four point pusher, however without suitable control over its homogeneity.  

The differences between the single point and the four points pushers can be understood and quantified using finite element analysis. The general setup configurations for the two different pushers are shown in  Figs.\ref{comsol}(a) and (b). Zooming into the $2 \times 2$ mm$^2$ central region of the Nitinol cross shows a much wider region with uniform in-plane strain with the four point pusher (Fig.\ref{comsol}(c)) than with the single point pusher (Fig.\ref{comsol}(d)). The color scale in these two panels represents the relative strain difference between the orthogonal directions $(\epsilon_{xx}-\epsilon_{yy})/\epsilon_{xx}(0)$, where $\epsilon_{xx}(0)=\epsilon_{yy}(0)$ is the strain at the center of the cross.

\section{CONCLUSION}

We have demonstrated a unique device capable of applying a uniform in-plane biaxial strain in excess of 1.1$\%$ to low dimensional single crystals and small devices. Key design features to achieve such uniform strain in a small device are the shapes of the bending substrate and pusher. The device is also capable of generating non-uniform biaxial strain, although the calibration of the strain along the two orthogonal directions remains an open issue. The device is magnetic field, low temperature and ultra-high vacuum compatible. It can be scaled to various experimental setups, including local probes and angular resolved photoemission. The Nitinol plates are capable of sustaining even larger strain than the 2.3$\%$ used here, the limiting factor in our current setup being the range of the in-situ actuated pusher. 

\section*{ACKNOWLEDGEMENTS}
This work was supported by the Swiss National Science Foundation (Division II Grant No. 182652).
\bibliography{paper.bib}

\end{document}